\documentclass[aps, prd, reprint,showpacs,  preprintnumbers,amsmath,amssymb,superscriptaddress,  nofootinbib]{revtex4-1}

\usepackage{silence}
\usepackage{comment}

\usepackage{amssymb, amsmath, braket, nicefrac, graphicx} 
\usepackage[tight]{subfigure}
\usepackage{relsize}
\usepackage{lineno}
\usepackage{setspace}

\usepackage{ esint }
\usepackage{chngpage}
\usepackage{epsfig}
\usepackage{multirow}
\usepackage{rotating}
\usepackage{float}
\usepackage{color}
\usepackage{xcolor}
\usepackage{appendix}
\usepackage{ amstext }
\usepackage{ gensymb }
\usepackage{ textcomp }
\usepackage{ wasysym }

\usepackage{upgreek}

\usepackage{pstricks}

\newcommand{\numu}{\mbox{$\nu_{\mu}$}}                   
\newcommand{\nue}{\mbox{$\nu_{e}$}}                      
\newcommand{\nutau}{\mbox{$\nu_{\tau}$}}                 

\newcommand{\anue}{\ensuremath{\bar{\nu}_{e}}}

\newcommand{\anumu}{\ensuremath{\bar{\nu}_{\mu}}}

\newcommand{\simgt}{\,\hbox{\lower0.6ex\hbox{$\sim$}\llap{\raise0.6ex\hbox{$>$}}}\,}
\newcommand{\simlt}{\,\hbox{\lower0.6ex\hbox{$\sim$}\llap{\raise0.6ex\hbox{$<$}}}\,}

\definecolor{maroon}{RGB}{162,10,10}

\usepackage{hyperref}
\usepackage[all]{hypcap}
\hypersetup{
    colorlinks,
    citecolor=black,
    filecolor=black,
    linkcolor=black,
    urlcolor=black
}

\makeatletter

\renewenvironment{figure}
  {\def\@captype{figure}}
  {}
\makeatother
\begin{document}

\title{\bf Decoherence, matter effect, \texorpdfstring{$\nu$}{neutrino} hierarchy signature in long-baseline experiments}
\newcommand{\Tufts}{Physics Department, Tufts University, Medford, Massachusetts 02155, USA}
\newcommand{\APC}{APC, Universit\'{e} Paris Diderot, CNRS/IN2P3, Sorbonne Paris Cit\'{e}, F-75205 Paris, France}
\newcommand{\emailcoelho}{jcoelho@apc.in2p3.fr}
\newcommand{\emailmann}{anthony.mann@tufts.edu}

\author{Jo\~{a}o A. B. Coelho}\thanks{\emailcoelho}\affiliation{\Tufts} \affiliation{\APC}
\author{W. Anthony Mann}\thanks{\emailmann}\affiliation{\Tufts}


\begin{abstract}
Environmental decoherence of oscillating neutrinos of strength $\Gamma = (2.3 \pm 1.1) \times 10^{-23}$ GeV
can explain how maximal $\theta_{23}$ mixing observed at 295 km by T2K appears to be non-maximal at
longer baselines.   As shown recently by R. Oliveira, the MSW matter effect for neutrinos is altered by decoherence:
In normal (inverted) mass hierarchy, a resonant enhancement of $\numu (\anumu) \rightarrow \nue (\anue)$
occurs for $6 < E_{\nu} < 20$ GeV.   Thus decoherence at the rated strength may be detectable 
as an excess of charged-current $\nue$ events in the full $\numu$ exposures of MINOS+ and OPERA.
\end{abstract}

\maketitle

\section{Introduction}
\subsection{\texorpdfstring{$\nu$}{Neutrino} decoherence in long-baseline experiments}

Neutrino flavor oscillations are the consequence of quantum mechanical mixing between mass and flavor eigenstates.
The mixing is generally regarded to be fully described by the PMNS matrix which is conventionally
parameterized using three mixing angles, $\theta_{12}$, $\theta_{13}$, $\theta_{23}$, and a CP violating phase, 
$\delta_{CP}$~\cite{PMNS-matrix}.    For nearly two decades neutrino oscillation experiments have indicated 
the $\theta_{23}$ angle, which characterizes flavor mixing of the `atmospheric' $\mu-\tau$ flavor sector, is compatible
with the maximal value of $45^{o}$ (i.e. $\sin^2\theta_{23} = 0.5$).    A close-to-maximal value for $\theta_{23}$ 
is very intriguing because it implies that the $\nu_3$ mass eigenstate is comprised of $\numu$ and $\nutau$ flavors in nearly equal amounts.
Recently however, there is an indication from the NOvA long-baseline neutrino oscillation experiment that
in $\numu$ disappearance oscillations observed at 810 km, flavor mixing in the atmospheric sector 
deviates from maximal mixing.   This indication for non-maximal mixing in NOvA data is reported to be 
at the level of 2.6\,$\sigma$~\cite{NOvA-nonmax}.   In a recent global analysis maximal mixing is disfavored at the 2\,$\sigma$ level~\cite{NuFit}.
The NOvA result is in tension with the 295-km baseline measurements of T2K~\cite{T2K-2017}, and it undermines the idea that
the PMNS lepton-flavor mixing matrix harbors an exact $\mu - \tau$ flavor symmetry~\cite{Xing-Zhao}.  In a recent paper
the authors proposed that neutrino decoherence that is environmentally-induced can help to alleviate this tension.   It is 
proposed that $\theta_{23}$ may in fact be nearly maximal and that the apparent tension can be lifted by neutrino decoherence
of coupling strength, $\Gamma$, in a specified range whose influence becomes more pronounced
with increasing oscillation baselines~\cite{Decoh-scenario}.  
Previous studies of neutrino decoherence have considered various integer power-law forms for the decoherence parameter:
$\Gamma = \Gamma_{0} \cdot (\frac{E_{\nu}}{[GeV]})^{n}$~\cite{Lisi-2000, Fogli-2007, Oliveira-2014}.   For experiments
that have similar ratios of baseline to neutrino energy, $L/E_{\nu}$, such as T2K and NOvA however, the tension cannot be
alleviated by a negative-integer power-law model.    On the other hand, decoherence effects that ensue with a positive-integer
power law form are constrained by atmospheric neutrino data~\cite{Lisi-2000}.    The model considered by this work assumes 
decoherence to be characterized by a coupling parameter that is energy-independent 
and of strength $\Gamma = (2.3 \pm 1.1) \times 10^{-23}$ GeV~\cite{Decoh-scenario}.   Decoherence of this form
 introduces a small, exponential damping of neutrino oscillations
that mimics the effect of non-maximal $\theta_{23}$ mixing when interpreted by conventional 3-flavor neutrino-oscillation analyses.

Decoherence is observed in a variety of quantum systems that are ``open" to environmental influences, and the phenomenology for
describing decoherence effects in these systems is well developed~\cite{Weinberg-LQM}.  
For evolving neutrino states, the pervasive environment might originate in couplings 
to new physics beyond the standard electroweak model, e.g. perturbations arising from 
spacetime itself and its Planck-scale dynamics~\cite{Spacetime-1995, Lisi-2000, Barenboim-2006}.

To be clear, environmentally induced decoherence is to be distinguished 
from neutrino wave-packet decoherence, a quantum wave effect 
that one may expect to occur based on known physics~\cite{Chan-2016, Daya-Bay-2016}.   
While wave-packet decoherence also introduces
exponential damping factors that multiply the oscillatory terms in the oscillation transition probabilities, 
the damping depends strongly on the neutrino energy as well as on the baseline.    
Consequently wave-packet decoherence is not viable as an effect that can account for the
emergence of apparent nonmaximal neutrino mixing with a longer baseline~\cite{Decoh-scenario}.
 
 The purpose of this paper is to highlight a remarkable prediction concerning neutrino decoherence and to show,
 with numerical analyses, that experimental detection of a specific decoherence signature is currently within the
 reach of the MINOS+ and OPERA accelerator-based long-baseline experiments.

 \subsection{\texorpdfstring{$\nu$}{Neutrino} decoherence alters the matter effect}
 
Neutrino environmental decoherence, if present, may conspire with the terrestrial MSW matter effect~\cite{ref:MSW} to give a small resonant peak 
in $\nue$-flavor appearance oscillations in the neutrino energy range $6 < E_{\nu} < 20$ GeV.    As will be elaborated in this work,
an enhancement in the charged current (CC) $\nue$-flavor rate is predicted which is large enough to be observable 
by current-generation accelerator-based long-baseline experiments.
That is, neutrino decoherence would enable the MSW matter effect 
to provide the same hierarchy signature that, sans decoherence, 
could only appear with baselines that predominantly involve mantle traversal. 
Observation of this enhancement in $\numu (\anumu) \rightarrow \nue (\anue)$ 
oscillations would, at once, be strong evidence for neutrino environmental decoherence 
and for the normal (inverted) mass hierarchy.

The unusual interplay between neutrino decoherence and the MSW matter effect 
was first uncovered by Roberto Oliveira, who
considered its implications for oscillation measurements at DUNE~\cite{Oliveira-2016}.    
Unfortunately, as Oliveira acknowledged,
the resonant enhancement is likely beyond the reach of the neutrino beams 
envisaged for DUNE, whose flux spectral peaks are in the range
2.5 to $\sim 3.0$ GeV~\cite{Strait-2015} and whose intensities 
by $E_{\nu} \sim 8$ GeV are down by 1.5 orders-of-magnitude.
The situation is even less favorable with the off-axis NuMI beam to NOvA; 
the beam, by design, is narrow-band with the spectral peak located at 2 GeV.
With the medium-energy exposure of MINOS+ to the on-axis NuMI beam at Fermilab 
however, the possibility exists for detection of this 
decoherence - matter effect ``conspiracy".    

\subsection{Overview}
We begin with a heuristic derivation of $\numu \rightarrow \nue$ appearance oscillations with inclusion of decoherence.   
We use minimalist phenomenology to show
how decoherence enters into the familiar leading-order oscillation formulas for 3-flavor oscillations with matter effects and 
how it modifies the relationship between the oscillation amplitude and the oscillation phase.  We arrive at the same decoherence parametrization presented in our previous work and its forms, albeit approximate, exhibit the essential physics.  More accurate, albeit complicated, analytic forms are available in the literature~\cite{Oliveira-2016}, and more considered justifications concerning the reduction of coupling parameters can be found in~\cite{Decoh-scenario} and elsewhere~\cite{Oliveira-2016, Gago-2002, Oliveira-EurPhys-2010, Oliveira-2013, Guzzo-2016, Gomes-2017}.

Precise numerical methods are then used to calculate the probabilities for the $\numu (\anumu) \rightarrow \nue (\anue)$ transition in current long-baseline
experiments.  The latter calculations represent a full three-flavor plus matter effects treatment, taking into account the solar-scale mixing and CP-violation contributions that are neglected in our heuristic analytic result.
With the oscillation-with-decoherence probabilities in hand, we consider the prospects for detection of the decoherence-driven enhancement in $\numu (\anumu) \rightarrow \nue (\anue)$ predicted to occur in the vicinity of the terrestrial MSW resonance at $E_{\nu} \sim 12\mbox{ GeV} \left(\frac{\Delta{m^2_{31}}}{2.52\times10^{-3}\mbox{\scriptsize\ eV}^2}\right) \left(\frac{2.75\mbox{\scriptsize\ g/cm}^3}{\rho}\right)$.  OPERA and MINOS+ are experiments whose beams give non-negligible event rates in their far detectors in the energy region $6 < E_{\nu} < 20$ GeV.    Preliminary results from the OPERA experiment's search for $\nue$ appearance do not show an excess of events, however the event statistics are low in the relevant region of $E_{\nu}$.   The MINOS+ experiment will isolate a candidate sample with higher statistics but with much larger backgrounds, assuming that the experiment analyzes its full medium-energy exposure of $10^{21}$ POT into $\numu$ neutrinos of the NuMI beam at Fermilab.    The search by MINOS+ can be augmented by examining events recorded by the earlier MINOS exposures in the low-energy NuMI beam.

The neutrino fluxes and/or flux-times-cross-section event rates per exposure protons-on-target (POT) have been reported by MINOS, MINOS+, and OPERA at conferences, as have their event rates after selection for multi-GeV CC $\nue$-like events.   Based on this information we have assembled rough predictions for the $\nue$ rate excess that these experiments should see if indeed decoherence is operative and neutrino masses follow the normal mass hierarchy. 
We find that the preliminary data analyses presented by OPERA and MINOS+ point towards opposite conclusions.
We conclude by urging for more focussed experimental effort to shed light on the remarkable possibility that environmental decoherence engages with the MSW matter effect to open a window onto the ordering of neutrino masses.

\section{Decoherence unlocks matter effect in \texorpdfstring{$\numu \rightarrow \nue$}{numu to nue} oscillation}

That environmental decoherence may conspire to save the MSW matter effect from self-imposed irrelevance in current 
long-baseline oscillation experiments is an important result that is generally not appreciated.   
The mechanism by which this happens, however, can be readilly elicited 
by considering analytic forms for the $\numu \rightarrow \nue$ transition in matter in the presence of decoherence.

For neutrino states considered as a closed system, time evolution is governed by the total Hamiltonian $\hat{H} = \hat{H}_{osc} + \hat{V}_{matter}$.    Since $\hat{H}$ is hermitian, there is a basis spanned by neutrino effective mass states wherein $\hat{H}$ is diagonal:
\begin{equation} 
\label{eq:H-in-effective-mass-basis}
\hat{H} = \textrm{diag}\,( \tilde{E}_{1}, \tilde{E}_{2}, \tilde{E}_{3} )_{\tilde{m}}
\end{equation}
For the elements we write $ \tilde{E}_{i} \equiv \tilde{m}_i^2 /2 E_{\nu}$, 
associating energy levels to effective neutrino masses in analogy to oscillations described in vacuum. 
The importance of differences in the squared effective masses can be seen upon re-phasing the Hamiltonian, 
an algebraic manipulation leading to removal of a term proportional to $\mathbb{\hat{I}}$ which merely contributes 
an overall phase to the oscillation amplitudes.  The re-phased Hamiltonian is 
\begin{equation} \label{eq:re-phased-H-1}
\hat{H} = \frac{1}{2 E_{\nu}} \textrm{diag}( 0 , \Delta \tilde{m}_{21}^2, \Delta \tilde{m}_{31}^2 )_{\tilde{m}}.
\end{equation}
From low energy measurements, where matter effects are negligible, the neutrino masses are known to follow the pattern 
$|m_{2}^2 - m_{1}^2| \ll |m_{3}^2 - m_{1}^2|$, i.e. $|\Delta m_{31}^2| \simeq  |\Delta m_{32}^2| \gg |\Delta m_{21}^2|$.
Furthermore, for the baselines, $L$,  and neutrino energies, $E_{\nu}$ that
characterize the accelerator-based long-baseline experiments, $\Delta m_{21}^2 L/ 4E_{\nu} \ll 1$.   
The contribution from $\Delta m_{21}^2$ in vacuum may thus be neglected. Additionally we neglect the Dirac CP phase, $\delta_{CP} \rightarrow 0$. (Below we show complex conjugations of mixing matrix elements, even though they have no effect in our formulas.)

For accelerator neutrinos evolving/propagating sans decoherence, time in natural units equals the propagation distance, $L$, and the oscillatory phase is conveniently expressed as $\Delta\,\cdot\,L$ using the form $\Delta \equiv (\Delta m_{31}^2/4 E_{\nu})$.   Under these assumptions, the re-phased Hamiltonian can be easily solved analytically yielding
\begin{equation} \label{eq:diagonalized-H}
\hat{H} \simeq 2 \Delta \, \textrm{diag}( 0, \tilde{\xi}_{-}, \tilde{\xi}_{+})_{\tilde{m}},
\end{equation}
\noindent where the matter-effective phases that describe neutrino oscillations 
in matter of constant density~\cite{Freund-PRD-2001, Benatti-2001} are
\begin{equation} \label{eq:matter-phases}
\begin{split}
&\tilde{\xi}_{\pm} = \frac{1}{2}\left[ \hat{A} + 1 \pm \tilde{\xi}_0 \right], ~~\textrm{and}\\
&\\
&\tilde{\xi}_0 = \sqrt{\sin^2 2\theta_{13} + (\cos 2 \theta_{13}-\hat{A})^2}.  
\end{split}
\end{equation}
Here, $\hat{A}$ is the terrestrial matter potential,
\begin{equation} 
\label{eq:matter-potential}
 \hat{A} =  \pm \frac{\sqrt{2} \, G_F \, n_e}{2\Delta}
\end{equation}
\noindent where $G_F$ is the Fermi constant, $n_e$ is the electron density 
in matter, and the $+$ ($-$) sign applies to neutrinos (antineutrinos).
From equations \eqref{eq:re-phased-H-1} and \eqref{eq:diagonalized-H}, 
we can then derive the following associations:
\begin{equation} \label{eq:effective-deltas}
\begin{split}
&\tilde{\Delta}_{21} \approx \Delta \, \tilde{\xi}_{-} ,\\
&\tilde{\Delta}_{31} \approx \Delta \, \tilde{\xi}_{+} ,\\
&\tilde{\Delta}_{32} \approx \Delta \, \tilde{\xi}_{0} .
\end{split}
\end{equation}

\noindent
In the absence of decoherence, the time evolution of neutrino states is governed 
by the effective Schr\"odinger wave equation or equivalently, 
using density matrices for pure states, by the von Neumann equation.   

With environmental decoherence operative however, the neutrino states no longer comprise an isolated system.  Their coupling to the environment has consequences for time evolution which are embodied by the Lindblad master equation.   The presence of weakly perturbative dynamics is parameterized by a ``dissipator" term, $\mathcal{D}[\hat{\rho}(t)]$, added to the von Neumann equation~\cite{Gorini-1978, Lindblad}.   We undertake to solve the Lindblad equation, starting with the Hamiltonian of Eq.~\eqref{eq:H-in-effective-mass-basis} and the density matrix $\rho_{\tilde{m}}(t)$ expressed in the effective mass basis:
\begin{equation} 
\label{eq:Lindblad-mass-basis}
    \frac{d}{dt}\hat{\rho}_{\tilde{m}}(t) = - i [\hat{H} , \hat{\rho}_{\tilde{m}}(t) ] - \mathcal{D}[\hat{\rho}_{\tilde{m}}(t)].
\end{equation}

The general form of the dissipator is determined by the requirement of {\it complete positivity}~\cite{Lindblad, Benatti-1997}.   It is constructed using a set of $N^2 -1$ operators, $\hat{D}_{n}$, where $N$ is the dimension of the 
Hilbert space of interest.  So, for three-flavor neutrino oscillations $N = 3$ and the $\hat{D}_{n}$ 
can, for example, be expressed as linear combinations of the Gell-Mann matrices. 
\begin{equation} 
\label{eq:Lindblad-dissipator}
\begin{split}
&\mathcal{D}[\hat{\rho}_{\tilde{m}}(t)] = \\   
 &\sum_{n = 1}^8 \left( \{ \hat{\rho}_{\tilde{m}}(t) , \hat{D}_{n}^{\dagger}\hat{D}_{n} \}_{+} - 2 \hat{D}_{n}  \hat{\rho}_{\tilde{m}}(t) \hat{D}_{n}^{\dagger} \right) .
\end{split}
\end{equation}
Constraints are imposed on the $\hat{D}_{n}$ arising from mathematical considerations 
and from the laws of thermodynamics.  Specifically, it is assumed that the von Neumann entropy, 
$S = - {Tr} (\hat{\rho} \,\textrm{ln}\hat{\rho} )$, increases with time and this is enforced 
by requiring the $\hat{D}_{n}$ to be hermitian.  In addition, conservation of the average energy of the system, calculated as ${Tr}(\hat{\rho} \,\hat{H})$, 
is assured by requiring $[ \hat{D}_{n}$, $\hat{H}] = 0$ for all $\hat{D}_{n}$.   Thus 
average conservation of energy implies that the $\hat{D}_{n}$ and $\hat{H}$ can be simultaneously diagonalized.

Our approach to a solution follows the phenomenological approach presented 
in Ref.~\cite{Farzan-2008}, however here we assume that the basis in which the simultaneous diagonalization happens, i.e. the basis in which the conservation of energy is defined, is not the neutrino mass basis, but rather the effective mass basis:
\begin{equation}
\label{eq:D-in-effective-mass-basis}
\hat{D}_{n} = \textrm{diag}\,( d_{n,1}, d_{n,2}, d_{n,3} )_{\tilde{m}}  .
\end{equation}
Considering the nth term in the dissipator as indicated by the right-hand side of Eq.~\eqref{eq:Lindblad-dissipator}, 
the anticommutator of $\hat{\rho}_{\tilde{m}}(t)$ with the product 
$\hat{D}_{n}^{\dagger} \hat{D}_{n} = \textrm{diag}\,(d_{n,1}^2, d_{n,2}^2, d_{n,3}^2)$ gives
\begin{equation} \label{eq:anticommutator}
 \begin{split}
 & \{ \hat{\rho}_{\tilde{m}}(t) , \hat{D}_{n}^{\dagger}\hat{D}_{n} \}_{+} = \\
    &\left( \begin{array}{ccc}
                      2 d_{n,1}^2 \,\rho_{11}  & (d_{n,1}^2 + d_{n,2}^2) \, \rho_{12} & (d_{n,1}^2 + d_{n,3}^2) \, \rho_{13} \\
                       (d_{n,1}^2 + d_{n,2}^2) \, \rho_{21}  &  2 d_{n,2}^2 \,\rho_{22}  & (d_{n,2}^2 + d_{n,3}^2) \, \rho_{23}  \\
       (d_{n,1}^2 + d_{n,3}^2) \, \rho_{31}  & (d_{n,2}^2 + d_{n,3}^2) \, \rho_{32}   & 2 d_{n,3}^2 \,\rho_{33}  \\
    \end{array} \right)
\end{split} 
\end{equation}
Upon addition of the second part of the nth term, $- 2 \hat{D}_{n} \, \hat{\rho}_{\tilde{m}}(t) \,\hat{D}_{n}^{\dagger}$,
with the anticommutator, the diagonal elements of \eqref{eq:anticommutator} are canceled and the off-diagonal
elements become amenable to completion-of-squares.  Summing over all eight terms 
gives constant, positive-valued decoherence couplings in the 
off-diagonal elements that we write as
\begin{equation} 
\label{eq:dissipation-parameters}
\Gamma_{ij} = \Gamma_{ji} = \sum_{n = 1,8} (d_{n,i}-d_{n,j})^2 .
\end{equation}
Thus the dissipator term in effective mass basis is 
\begin{equation} \label{eq:dissipator-matrix}
 \begin{split}
&- \mathcal{D}[\hat{\rho}_{\tilde{m}}(t)] =    
    \left( \begin{array}{ccc}
                        0  & - \Gamma_{21} \, \rho_{12}(t) &  - \Gamma_{31} \, \rho_{13}(t) \\
                        - \Gamma_{21} \, \rho_{21}(t)  &  0  &  - \Gamma_{32} \, \rho_{23}(t)  \\
         - \Gamma_{31} \,  \rho_{31}(t) &  - \Gamma_{32} \, \rho_{32}(t)   & 0  \\
    \end{array}\right) .
\end{split} 
\end{equation}
Adding matrix \eqref{eq:dissipator-matrix} to the von Neumann commutator 
in the right-hand side of Eq.~\eqref{eq:Lindblad-mass-basis} gives the 
Lindblad equation in matrix form.    First integration of each element yields
\begin{equation} \label{eq:solution-form-1}
 \begin{split}
  \hat{\rho}_{\tilde{m}}(t) & = \\
   & \left( \begin{array}{ccc}
                         \rho_{11}(0) &   \rho_{12}(0) \, e^{- \eta_{21}^*t}  & \rho_{13}(0) \, e^{- \eta_{31}^*t}   \\
                          \rho_{21}(0)\, e^{- \eta_{21} t} &  \rho_{22}(0)  & \rho_{23}(0)  e^{- \eta_{32}^*t}     \\
                        \rho_{31}(0) e^{- \eta_{31}t}  & \rho_{32}(0) e^{- \eta_{32}t}   &  \rho_{33}(0) \\
    \end{array} \right).
\end{split} 
\end{equation}
where we have introduced the notation $\eta_{ij} \equiv (\Gamma_{ij} + i 2 \tilde{\Delta}_{ij})$.

For our purposes it suffices to assume that only a single $\hat{D}_{n}$ operator is active:  $\hat{D}_{n} = \textrm{diag}( d_1, d_2, d_3)$.
Its eigenvalues are real, positive-valued, quantities with dimension of energy, 
which we assume to be related in a way that mimics the neutrino mass pattern
$|m_{2}^2 - m_{1}^2| \ll |m_{3}^2 - m_{1}^2|$, namely $ (d_{2} - d_{1})^2 \ll (d_{3} - d_{1})^2$. 
Additionally, in order to focus on the dominant decoherence effect, we 
assume that $(d_{2} - d_{1})^2 \simeq 0$ and can be ignored.
The latter assumption together with Eq.~\eqref{eq:dissipation-parameters} 
reduces the set of decoherence couplings to a single parameter, $\Gamma$: 
\begin{equation} 
\label{eq:decoherence-parameters}
\Gamma_{21} = 0 ~~\textrm{and}  ~~  \Gamma \equiv \Gamma_{31} = \Gamma_{32}.          
\end{equation}

The initial neutrino state $ \ket{\numu}$ is represented in the effective mass basis 
by the density matrix $\rho_{\tilde{m}}^{(\mu)}(0)$
whose elements are 
\begin{equation} 
\label{eq:initial-state-rho}
\hat{\rho}_{\tilde{m}}^{(\mu)}(0) =  \tilde{U}^{T} \cdot \hat{\Pi}^{\numu} \cdot  \tilde{U}^{*}  ~
\Leftrightarrow ~ \left[ \rho_{\tilde{m}}^{(\mu)}(0) \right]_{ij} = \tilde{U}_{\mu i}\tilde{U}_{\mu j}^{*}.
\end{equation}
Here we refer to the state projector in neutrino flavor basis  
$\hat{\Pi}^{\numu} \equiv \ket{\numu}\bra{\numu} \equiv \textrm{diag}(0, 1, 0)_{\alpha}$,
and to the PMNS mixing matrix with matter-effective elements:
\begin{equation} \label{eq:mixing-matrix-elements}
   \tilde{U} = 
  \left(\begin{array}{ccc}
    \tilde{U}_{e1} & \tilde{U}_{e2} & \tilde{U}_{e3} \\
    \tilde{U}_{\mu 1} & \tilde{U}_{\mu 2} & \tilde{U}_{\mu 3} \\
     \tilde{U}_{\tau 1} & \tilde{U}_{\tau 2} & \tilde{U}_{\tau 3}
  \end{array}\right).
\end{equation}
The unusual-looking form of the unitary transformation indicated in Eq.~\eqref{eq:initial-state-rho} reflects the 
convention $\ket{\nu(0)} = \ket{\nu_{\alpha}} = U^{*}_{\alpha j}\ket{\nu_j}$. 

Insertion of the initial-state elements of Eq.~\eqref{eq:initial-state-rho} into Eq.~\eqref{eq:solution-form-1} yields the time-evolved density matrix:
\begin{equation} \label{eq:evolved-density-matrix-2}
 \begin{split}
 \hat{\rho}_{\tilde{m}}&^{(\mu)}(t=L) = \\
    &\left( \begin{array}{ccc}
       |\tilde{U}_{\mu 1}|^2  & \tilde{U}_{\mu 1}\tilde{U}^*_{\mu 2}\, e^{- \eta_{21}^* L} & \tilde{U}_{\mu 1}\tilde{U}^*_{\mu 3} \, e^{- \eta_{31}^* L} \\
       \tilde{U}_{\mu 2}\tilde{U}^*_{\mu 1}\, e^{- \eta_{21} L} &  |\tilde{U}_{\mu 2}|^2   & \tilde{U}_{\mu 2}\tilde{U}^*_{\mu 3} \, e^{- \eta_{32}^* L} \\
       \tilde{U}_{\mu 3}\tilde{U}^*_{\mu 1}\, e^{- \eta_{31} L} & \tilde{U}_{\mu 3}\tilde{U}^*_{\mu 2}\, e^{- \eta_{32} L}  &  |\tilde{U}_{\mu 3}|^2  \\
    \end{array} \right).
\end{split} 
\end{equation}
To obtain the $\nue$ appearance probability at baseline $L$, we
transform the density matrix into the neutrino flavor basis $\{ \ket{\nu_{\alpha}}, \alpha = e, \mu, \tau \}$:
\begin{equation} 
\label{eq:into-flavor-basis}
 \rho_{\alpha}^{(\mu)}(t)  = \tilde{U}^* \cdot \rho_{\tilde{m}}^{(\mu)}(t) \cdot \tilde{U}^{T}.
\end{equation}
The state projector for the final state in $\numu \rightarrow \nue$ appearance oscillations, $\ket{\nue}$, is  
$\hat{\Pi}^{\nue} \equiv \ket{\nue}\bra{\nue} \equiv \textrm{diag}(1, 0, 0)_{\alpha}$, and the 
$\nue$ appearance probability is calculated as 
\begin{equation} 
\label{eq:probability-1}
 \mathcal{P}_{(\nu_\mu \rightarrow \nu_e)} = Tr\left[ \hat{\Pi}^{\nue} \cdot \rho_{\alpha}^{(\mu)}(t) \right]  = \bra{\nue}  \rho_{\alpha}^{(\mu)}(t)  \ket{\nue},
\end{equation}
or equivalently,
\begin{equation} 
\label{eq:probability-3}
 \mathcal{P}_{(\nu_\mu \rightarrow \nu_e)} = \sum_{ i,j} \tilde{U}^*_{e i} \tilde{U}_{e j} \left[ \rho_{\tilde{m}}^{(\mu)}(t) \right]_{ij} .
\end{equation}
Due to the symmetry introduced by neglecting the solar scale mass-square splitting, 
it can be shown that, for neutrinos (antineutrinos), the effective solar mixing angle $\tilde{\theta}_{12} = \pi/2\ (0)$, i.e. $\tilde{U}_{e1} = 0$ ($\tilde{U}_{e2} = 0$). Writing out Eq.~\eqref{eq:probability-3} explicitly, and taking $U_{e 1} = 0$, the appearance probability becomes
\begin{equation} 
\label{eq:probability-final-1}
\begin{split}
 \mathcal{P}_{(\nu_\mu \rightarrow \nu_e)} & =  |\tilde{U}_{e 2}|^2 |\tilde{U}_{\mu 2}|^2 + |\tilde{U}_{e 3}|^2 |\tilde{U}_{\mu 3}|^2 \\ & +
2 \,\mathcal{R}e \{ \tilde{U}^*_{e 2} \tilde{U}_{e 3}\tilde{U}_{\mu 2}\tilde{U}^*_{\mu 3} e^{- (\Gamma - 2i \tilde{\Delta}_{32} L)} \} .\\  
 \end{split}
\end{equation}

To facilitate comparison with commonly used probability expressions, we 
re-express the mixing matrix amplitudes in terms of the matter-effective mixing angles.
With neglect of solar-scale mixing, the mixing-matrix elements are given by
\begin{equation} 
\label{eq:mixing-matrix-elements-angles}
\begin{split}
\tilde{U} =& ~U_{23}(\theta_{23}) \, \tilde{U}_{13}(\tilde{\theta}_{13}, \delta_{CP} = 0) \, U_{12}(\theta_{12} = \pi/2) \\
& = U_{23} \, \tilde{U}_{13} =
  \left(\begin{array}{ccc}
           0 &  \tilde{c}_{13}       & \tilde{s}_{13} \\
     -c_{23} & -\tilde{s}_{13}s_{23} & \tilde{c}_{13} s_{23} \\
      s_{23} & -\tilde{s}_{13}c_{23} & \tilde{c}_{13} c_{23}
  \end{array}\right).
\end{split}
\end{equation}
Evaluating the mixing amplitudes and using the identity $c_\theta s_\theta  = s_{2\theta}/2$, we obtain
\begin{equation} 
\label{eq:probability-final-3}
 \mathcal{P}_{(\nu_\mu \rightarrow \nu_e)}  =  \sin^2 2\tilde{\theta}_{13} \cdot \sin^2 \theta_{23} \cdot  \frac{1}{2}[ 1 - e^{- \Gamma L} \cos(2\tilde{\Delta}_{32} L)].
\end{equation}
The matter-effective mixing angle $\tilde{\theta}_{13}$ is given by
\begin{equation} 
\label{eq:probability-8}
\tan 2\tilde{\theta}_{13} \simeq \frac{\sin 2\theta_{13}}{\cos 2\theta_{13} - \hat{A}}.
\end{equation}

The antineutrino probability can be written in exactly the same form, with the single replacement $\tilde{\Delta}_{32}\rightarrow\tilde{\Delta}_{31}$.

Equation~\eqref{eq:probability-final-3} gives the leading-order transition probability for $\numu \rightarrow \nue$ oscillations in the presence of environmental decoherence.    It is readily seen that in the limit $\Gamma \rightarrow 0$, the exponential damping factor becomes unity and allows the oscillatory $\cos(2\tilde{\Delta}_{32} L)$ term to combine with the first term via the identity $( 1 - c_{2 \phi} ) = 2 s^2_{\phi}$ to give the well-known, conventional $\nue$ appearance probability~\cite{Cervera-NPB-2000, Lipari-PRD-2000},
\begin{equation} 
\label{eq:probability-7}
 \mathcal{P}_{(\nu_\mu \rightarrow \nu_e)}  =   \sin^2(2\tilde{\theta}_{13}) \cdot \sin^2\theta_{23} \cdot \sin^2(\tilde{\Delta}_{32} L) .
\end{equation}

Comparing the conventional result \eqref{eq:probability-7} to oscillations with decoherence in-play, Eq.~\eqref{eq:probability-final-3},
one sees that the decoherence damping factor prevents the oscillation $\cos(2\tilde{\Delta}_{32} L)$ from being rolled together with the first term to give the conventional oscillatory behavior, $\sin^2(\tilde{\Delta}_{32} L)$.   Referring to Eq.~\eqref{eq:probability-final-3},
the presence of damping permits oscillatory cosine term - in the vicinity of MSW - to swing sufficiently 
far below 1.0 to give an enhancement in $\nue$ appearance.

These points are readily illustrated with numerical considerations; we use the OPERA/MINOS baseline of $\sim$730 km as representative.
The MSW resonance occurs in the vicinity of $E_{\nu} \sim 12$ GeV.   At the resonance, $(\cos 2 \theta_{13}-\hat{A}) \approx 0$,
$\sin^2 2\tilde{\theta}_{13} \rightarrow 1$, and $\tilde{\Delta}_{32} \rightarrow \Delta \cdot \sin 2\theta_{13}$.
Taking $\theta_{13} = 8.5^{\circ}$ as representative of the world-average value, we have $\sin 2\theta_{13} = \sin(17^{\circ}) \simeq 0.3$.   
Consequently the conventional oscillation phase at the value of $E_{\nu}$ that corresponds to the first oscillation maximum at 730 km, namely 1.5 GeV, gives $\sin^{2}(\frac{\pi}{2} \cdot \frac{1.5}{12} \cdot 0.3) \simeq \sin^{2}(\frac{\pi}{2} \cdot 0.04) \simeq 0.003.$   Thus 
the oscillation phase shuts down the probability for $\nue$ appearance oscillations, even though the effective mixing angles are at maximal strength.

But with decoherence $\Gamma = 2.3 \times 10^{-23}$ GeV in play, the oscillation phase exerts influence in a different way.
At the MSW resonance,  Eq.~\eqref{eq:probability-final-3} becomes
\begin{equation} \label{eq:minos_prob-4}
  \mathcal{P}_{(\nu_\mu \rightarrow \nu_e)}
   \simeq 1\cdot \frac{1}{2} \cdot  \frac{1}{2}\cdot [ 1 - e^{ - \Gamma L} \cdot\cos \left(\pi \cdot 0.04 \right)] .
\end{equation}
The cosine function is still close to 1.0 but is now damped by decoherence, consequently the MSW effect is still
rendered to be small, but not zero:  $\mathcal{P}_{(\nu_\mu \rightarrow \nu_e)} \simeq 0.022$.

\section{Numerical calculations}

The oscillation probabilities used in our analysis are derived from an exact formulation of Eq.~\eqref{eq:probability-3}, where the effective mixing elements of the PMNS matrix and the effective mass-square splittings were computed numerically through the diagonalization of the matter perturbed Hamiltonian.
A full three flavor treatment was implemented and the following values of the neutrino oscillation parameters were used~\cite{ref:global-fit}: $\sin^2\theta_{12}=0.306$, $\sin^2\theta_{13}=0.02166$, $\delta_{CP}=1.45\pi$, $\Delta{m^2_{21}}=7.50\times10^{-5}$\,eV$^2$, and $\Delta{m^2_{31}}=+2.524\times10^{-3}$\,eV$^2$. In our decoherence scenario~\cite{Decoh-scenario}, the $\theta_{23}$ angle is assumed to be maximal, i.e. $\sin^2\theta_{23}=0.5$. The density of matter is taken to be constant at a value of $\rho=2.8$\,g/cm$^3$.

\begin{figure}
\begin{adjustwidth}{0.75in}{0.75in}
\centering
\resizebox{\columnwidth}{!}{\includegraphics[angle=0, trim = 0mm 0mm 0mm 0mm, clip]{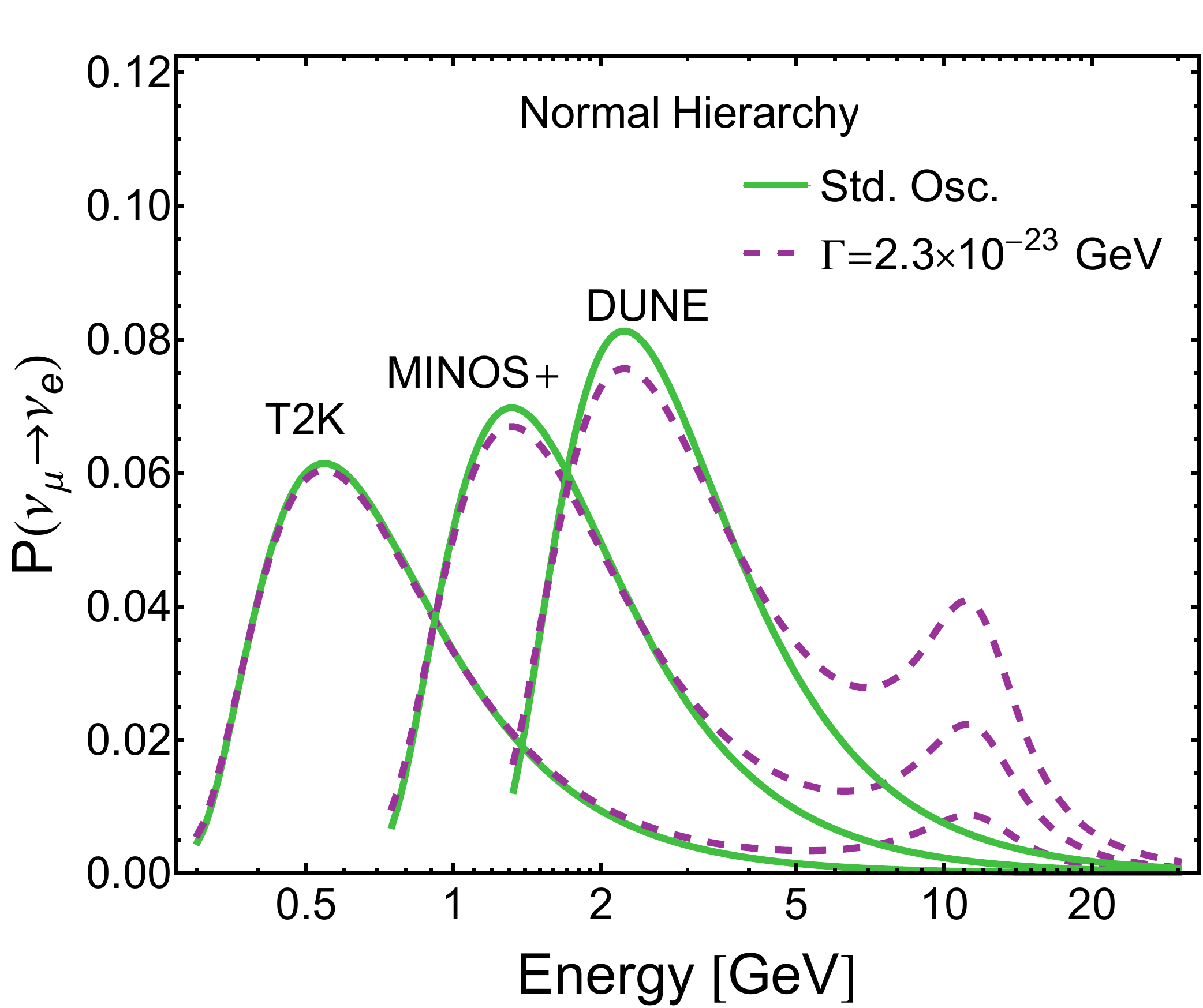}}
\end{adjustwidth}
\caption{Probability versus $E_{\nu}$ for $\numu \rightarrow \nue$ oscillations with environmental decoherence of strength $\Gamma$(dashed curve) for current long-baseline experiments, assuming the normal mass hierarchy. 
In the case that the effective energy basis is the eigenbasis for the environmental ``measurement", decoherence conspires with the terrestrial MSW matter effect to give a small $\nue$ appearance maximum at 12 GeV.}
\label{fig:Fig01}
\end{figure}

Results from numerical calculation of probabilities for the $\numu  \rightarrow \nue $ transition are shown in Fig.~\ref{fig:Fig01}, for the baselines of T2K (295 km), MINOS+ (735 km), and DUNE (1300 km);  the probabilities shown are for the normal mass hierarchy case.   For each baseline, the probability for oscillation with decoherence (dashed curve) is displayed together with the expectation for conventional 3-flavor oscillation (solid curve).  The comparisons thus provided show decoherence to diminish the transition in the main oscillation peak while promoting
a secondary enhancement at higher $E_{\nu}$ in the vicinity of the MSW resonance.  
The strength of both trends is seen to grow as the observational baseline is extended.
Probabilities for the OPERA (730 km) and NOvA (810 km) baselines are, of course,
very nearly the same as those shown for MINOS+.

\section{Prospects for detection}

The current generation of accelerator-based, long-baseline neutrino oscillation experiments was optimized to measure the parameters that govern Standard-Model (SM) 3-flavor neutrino mixing using $\numu$-flavor disappearance and $\nue$-flavor appearance measurements.   Unfortunately those optimizations mitigate against beams with abundant fluxes in the vicinity of the terrestrial MSW matter effect at $E_{\nu} \sim 12$ GeV.   For example, T2K is matched to a muon-neutrino beam whose flux spectrum peaks at 0.6 GeV and becomes negligible above 2-3 GeV.   Similarly NOvA operates off-axis in the NuMI beam at Fermilab, an arrangement that provides a narrow-band flux peaked near 2 GeV.   Prior to data-taking by NOvA, the MINOS experiment, whose far detector lies on-axis with respect to the NuMI beam, obtained all of its exposures with NuMI operated in its low-energy mode.  This configuration  gave a wide-band spectrum with flux peak at $\sim 3$\,GeV, and MINOS used it to obtain a total exposure of $10.56 \times 10^{20}$ POT.

NOvA's off-axis low-energy beam is obtained by operating the NuMI beam line in its medium energy mode.   In that mode the on-axis $\numu$ beam is wide-band, with spectral peak at 7 GeV and with fluxes of useful intensity extending to $\geq$ 12 GeV.   When the NuMI beam switched to extended medium-energy running for NOvA, MINOS became the MINOS+ experiment, which subsequently recorded a total exposure of $10 \times 10^{20}$ POT in the on-axis, medium-energy NuMI beam.   It is the addition of this latter exposure which gives MINOS/MINOS+ some sensitivity for detecting $\numu \rightarrow \nue$ appearance oscillations in the vicinity of the terrestrial MSW resonance.

Using its large exposure the medium-energy NuMI beam, MINOS+ is capable of identifying  $\nue$-CC interactions in the region $6 < E_{\nu} < 12$\,GeV with an efficiency of 78\%.   The experiment is using this capability to search for evidence of $\numu \rightarrow \nue$ that may be driven by coupling to sterile neutrino(s).  Initial results from analysis of $2.97 \times 10^{20}$ POT were reported at Neutrino 2016~\cite{MINOS+-2016} and at ICHEP 2016~\cite{MINOS+_Bkgd_Prediction}.  For neutrino interactions in the 6-12 GeV range, the predicted number of candidate events CC-$\nue$ events assuming standard 3-flavor oscillations is 56.7 events.   MINOS+ observes 78 events, an excess of 2.3$\,\sigma$~\cite{MINOS+-2016}.

The situation is somewhat different with respect to the OPERA experiment, originally designed to measure $\numu \rightarrow \nu_{\tau}$ oscillations.   The experiment used the CERN CNGS $\numu$ beam directed towards the OPERA emulsion ``bricks" detector located 730 km away at the Gran Sasso Underground Laboratory.  The beam is wide-band, with a flux that has a maximum `flattop' extending from $\sim$ 8~GeV to 24~GeV, with $\langle E_{\nu} \rangle = 17$~GeV.   The average detection efficiency for charged-current $\nue$ events is $\sim 15 \pm 3 \%$ for $E_{\nu} < 10$~GeV, and $40 \pm 4 \%$ for $10 < E_{\nu} < 20$~GeV~\cite{OPERA-2013}.   The total exposure for the duration of the data-taking from 2008 to 2012 is $1.797 \times 10^{20}$~POT.   Preliminary results have been reported for the experiment's $\nue$ appearance search using the full data set~\cite{OPERA-2017}. 

Taking inspiration from these preliminary results, we have assembled predictions for the outcome of measurements from OPERA and MINOS+, where the latter would use the full MINOS+ exposure combined with the earlier MINOS $\numu$ exposures.  Our estimate builds upon distributions of charged-current (CC) event rate (e.g. the $\nu$ flux times CC-inclusive cross section vs. neutrino energy), which are available for the MINOS $\numu$ exposure~\cite{MINOS_Rate}, for the MINOS+ exposure~\cite{MINOS+_Rate}, and for OPERA exposure~\cite{OPERA-2013}.  Estimates of detection rates for $\nue$ signal and background processes have been extracted from predictions given by MINOS+~\cite{MINOS+_Bkgd_Prediction}, scaled to the appropriate POT exposure, and OPERA~\cite{OPERA-2017}. For MINOS+, the selection efficiency is assumed to be constant versus neutrino energy for all types of interaction, while for OPERA efficiencies are estimated for each bin of 10~GeV width and a linear interpolation is taken between bin centers. Neutral-current (NC) predictions are not affected by oscillations and are thus kept unchanged apart from the appropriate scaling with exposure.

\begin{figure}
\begin{adjustwidth}{0in}{0in}
\centering
\resizebox{\columnwidth}{!}{\includegraphics[angle=0, trim = 0mm 0mm 0mm 0mm, clip]{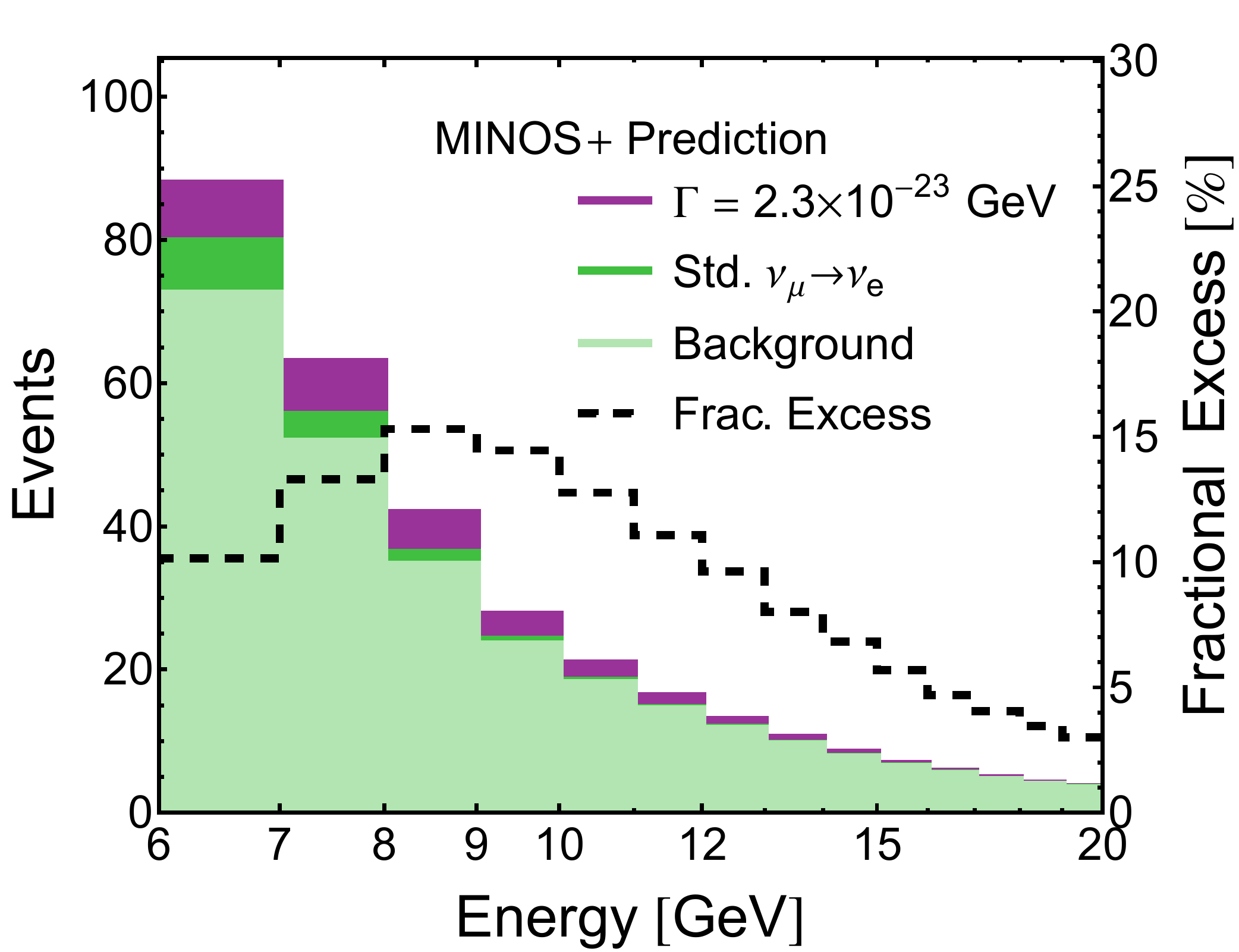}}
\end{adjustwidth}
\caption{Prediction for the excess of $\nue$ appearance events (uppermost histogram) that may occur in MINOS \& MINOS+ data for the case of normal mass hierarchy.  The enhanced $\nue$ rate arises from decoherence modification of the MSW resonance at $E_{\nu} = 12$ GeV.   The estimation is based upon publicly presented fluxes and detection efficiencies and assumes investigation of the full $(10.56 + 10) \times 10^{20}$ POT exposures of MINOS and MINOS+ to the low-energy and medium-energy NuMI beams.}
\label{fig:Fig02}
\end{figure}
\smallskip

Based on these estimates, our prediction for a MINOS \& MINOS+ analysis of a 6~GeV to 20~GeV search region in neutrino energy is presented in Fig.~\ref{fig:Fig02}.  The figure shows the predicted distribution of candidate $\nue$ events assuming that the decoherence-matter effect enhancement is operative and that the neutrino mass hierarchy is the normal hierarchy. The prediction 
indicates an excess of CC-$\nue$ candidates of $\sim 32$ events above an expectation of 282 events based on conventional 3-flavor oscillations. If a binned analysis is performed, the full MINOS \& MINOS+ data sample should be sensitive to this decoherence effect at $\sim 2\,\sigma$ statistical significance. 

In OPERA, the number of candidate CC-$\nue$ events with $0 < E_{\nu} < 20$~GeV predicted for the case of SM 3-flavor $\nu$ oscillations (including backgrounds and 3-flavor $\nue$ signal), is 7.0 events;  a total of 7 candidate events is observed.    However in the presence of the decoherence-matter effect enhancement, the rate of genuine $\nue$ appearance events would receive a boost: Instead of OPERA's estimate of 1.4 $\nue$-flavor events based upon conventional 3-flavor oscillations, the expected rate becomes $\sim$ 10.4 events.  The candidate sample to be expected is then 16.0 events, to be compared to 7 observed events.    Thus OPERA provides a test for the presence of the decoherence-matter effect enhancement in the case of normal hierarchy of $\sim 2.5\,\sigma$ statistical significance.

\section{Conclusions}

It is well-known that the terrestrial MSW matter effect, if subjected to experimental examination by mantle-crossing long-baseline $\numu/\anumu$ beams, can enable the hierarchy for neutrino mass states to be determined.  This paper highlights the related but new phenomenon identified by R.L.N. Oliveira: environmental decoherence of oscillating neutrino states could make the MSW $\numu \rightarrow \nue$ resonance and hierarchy discrimination accessible to existing experiments whose baselines are confined to the Earth's crust~\cite{Oliveira-2016}. We have identified OPERA and MINOS+ as two such experiments; both have reported preliminary measurements that have sensitivity to this phenomenon.  Results reported thus far however are tantalizingly ambiguous.   At the level of $\sim 2.5\,\sigma$ significance, OPERA observes no evidence for anomalous $\nue$ appearance for energies in the vicinity of the MSW resonance at 12 GeV.  Their result indicates that, if the mass hierarchy is normal then decoherence is absent or is, at best, associated with a smaller $\Gamma$ coupling than proposed by Ref.~\cite{Decoh-scenario}.   An opposite trend however -- an excess of $\nue$ charged-current event candidates in the interval 6\,GeV$< E_{\nu}<$12\,GeV -- is observed in MINOS+ data.   If the latter trend continues to hold in analysis of the experiment's full dataset, then decoherence with normal mass hierarchy offers a viable interpretation. This interpretation differs from the one that originally motivated the OPERA and MINOS+ investigations, namely that anomalous $\nue$ appearance may be the harbinger of coupling(s) to sterile neutrino(s).  To the extent that the current situation becomes more widely appreciated and further experimental investigation is encouraged -- this paper will have achieved its goal.


\section*{Acknowledgments} \vspace{-8pt}  
This work was supported by the United States Department of Energy under grant DE-SC0007866 and by
the IdEx program at Sorbonne Paris Cit\'{e} (ANR-11-IDEX-0005-02).


\end{document}